\documentclass[twocolumn,prl,aps,floats,showpacs,ansmath,amssymb]{revtex4}
\usepackage{times}
\usepackage{graphicx}
\usepackage{bm}

\begin{document}

\title{Correcting ray optics
at curved dielectric microresonator interfaces:\\ Phase-space
unification of Fresnel filtering and the Goos-H{\"a}nchen shift}

\author{Henning Schomerus}
\affiliation{Department of Physics, Lancaster University,
Lancaster, LA1 4YB, United Kingdom}
\author{Martina Hentschel}
\affiliation{Institut f{\"u}r Theoretische
Physik, Universit{\"a}t Regensburg, 93040 Regensburg, Germany }

\date{June 2006}

\begin{abstract}
We develop an amended ray optics description for reflection at the
curved dielectric interfaces of optical microresonators which
improves the agreement with wave optics by about one order of
magnitude. The corrections are separated into two contributions of
similar magnitude, corresponding to ray displacement in
independent quantum phase space directions, which can be
identified with Fresnel filtering and the Goos-H\"anchen shift,
respectively. Hence we unify two effects which only have been
studied separately in the past.
\end{abstract}
 \maketitle

Over the recent years it has become feasible to design optical
microresonators that confine photons by means of dielectric
interfaces into a small spatial region not larger than a few
micrometers \cite{YS93,VKLISLA98,QSTC86}.
 Two promising
lines of research are the amplification of photons by stimulated
emission in active media, which yields lasing action
\cite{YS93,VKLISLA98,QSTC86,bowtie,starofdavid,sangwookkim,morelasers,morelasers2,koreaner,japaner,lebental},
and the generation and trapping of single photons which can be
used as carriers of quantum information \cite{photons}. These
applications require integration of several components and
interfacing with electronics, which are best realized in
two-dimensional resonator geometries where the main in- and
out-coupling directions are confined to a plane, and can be
selected via the (asymmetric) resonator geometry. Furthermore,
because of the requirements of mode selection, these applications
favor microresonators of mesoscopic dimensions, with size
parameters $kL=O(100)-O(1000)$ (where $L$ is the linear size,
$k=2\pi/\lambda$ is the wavenumber and $\lambda$ is the
wavelength) which quickly puts these systems out of the reach of
numerical simulations. On the other hand, ray-optics predictions
of the intricate resonator modes
\cite{bowtie,sangwookkim,morelasers,nature,optlett_interf,pre_annbill,jan,schwefel}
can deviate substantially from experimental observations
\cite{starofdavid,japaner} and theoretical predictions
\cite{starofdavid,pre_annbill,jan,koreaner}.

The purpose of this paper is to develop an amended ray optics
(ARO) description which still  idealizes beams as rays, but
incorporates corrections of the origin and propagation direction
of the reflected ray. We identify these corrections by utilizing
quantum-phase space representations of the incident and reflected
beam \cite{epl_husimi} and relate them to the recently discovered
Fresnel filtering effect \cite{ff} and the long-known
Goos-H{\"a}nchen shift  \cite{ghs_entdeck}. These two effects have
only been discussed separately in the past (for applications to
microresonators see, e.g., Refs.\
\cite{koreaner,starofdavid,chowdhury,wir_pre}), and their
complementary nature  has not been realized. Moreover, their
uniform analysis for all angles of incidence is known to pose
considerable technical challenges
 \cite{ff,ghs_lotsch,ghs_lai,branchpoints}.
 In the phase-space representation, the Fresnel filtering and
Goos-H{\"a}nchen corrections are simply determined by the position
of maximal phase-space density. For the prototypical case of a
Gaussian beam reflected from a constantly curved dielectric
interface, we find that compared to conventional ray optics, the
resulting ARO improves the agreement of the far-field radiation
characteristics with wave optics by about one order of magnitude.

\begin{figure}
\includegraphics[width=0.48\columnwidth]{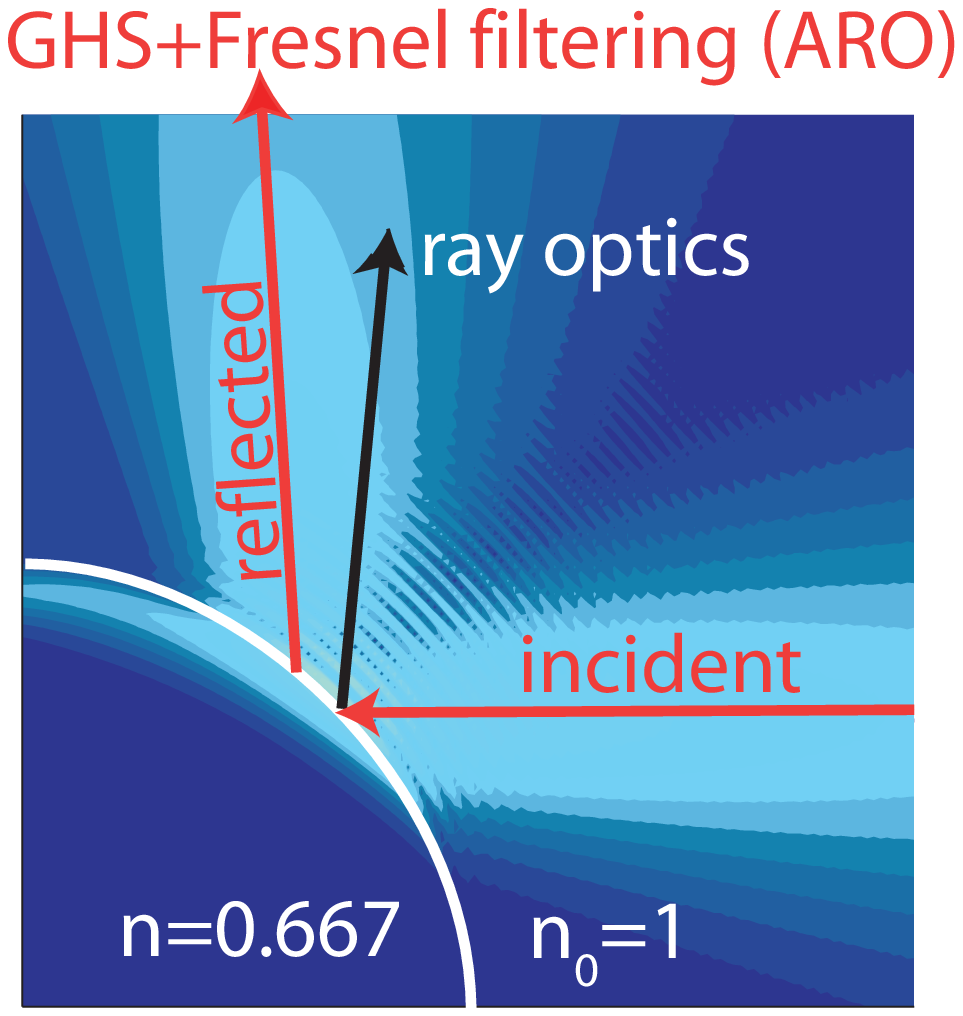}
\raisebox{.5cm}{\includegraphics[width=0.48\columnwidth]{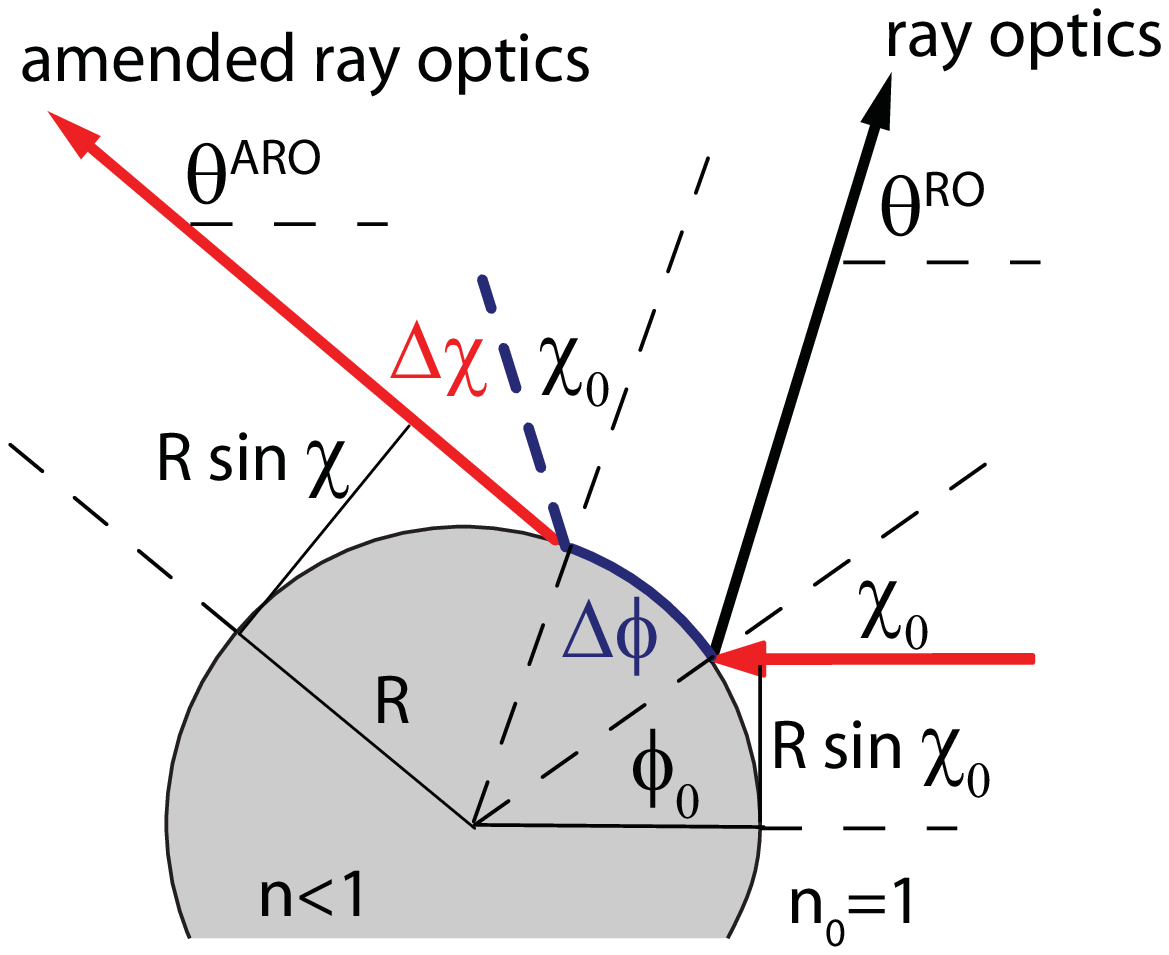}}
\caption{(Color online) Left panel: Gaussian beam reflected from a
curved dielectric interface ($kR=100$) separating regions of
refractive index $n_0=1$ and $n=0.667$. Light regions indicate
high wave intensity. The angle of incidence $\chi_0=42^\circ$ is
close to the critical angle $\chi'=41.8^\circ$. Conventional ray
optics predicts that the beam is specularly reflected at the point
of incidence. In this paper we use phase-space representations to
obtain a more accurate reflection law, which accounts for (i) the
Goos-H{\"a}nchen shift (GHS) $\Delta\phi$ of the reflection point
along the interface and (ii) the increase $\Delta \chi$ of the
reflection angle due to Fresnel filtering. Both effects change the
far-field radiation direction $\theta$ (see the right panel, which
exaggerates the corrections in order to clarify the notation). For
the parameters in the left panel, $\Delta\phi\approx 7^\circ$ and
$\Delta \chi\approx 1^\circ$ (see Fig.\ \ref{fig3}), resulting in
a corrected ray which nicely reproduces the observed radiation
direction.
  }\label{fig1}
\end{figure}

The plan of this paper is as follows. We first quantify the
corrections to conventional ray optics by phase-space
representations of the wave in the near field of the interface.
Then we incorporate these corrections into an ARO. Finally, we
test the ARO by its predictions for the far-field radiation
characteristics.

Conventional ray optics assumes that beams have well-defined
propagation directions and a precise point of impact on a sharp
dielectric interface, and predicts that an incident beam is
reflected specularly and locally at the interface \cite{BornWolf}.
In two dimensions, deviations from ray optics at curved interfaces
are apparent already at inspection of wave patterns such as shown
in the left panel of Fig.\ \ref{fig1}, where the incident beam
propagates from right to left. The wave pattern reveals that the
reflected beam originates from a
displaced position and propagates into a different direction than
predicted by ray optics.

We choose a coordinate system with origin at the center of the
circle of curvature (see the right panel of Fig.\ \ref{fig1}).
This circle has the same radius of curvature $R$ as the dielectric
interface and touches it tangentially at the point of incidence.
The incident  beam propagates into negative $x$ direction. For the
comparison of wave optics to ray optics it is convenient to
parameterize the rays by Birkhoff coordinates $(\phi,\sin\chi)$,
where $\phi$ parameterizes the polar angle of the ray's
intersection point with the interface, while $R\sin\chi$ is the
impact parameter of the ray, where $\chi$ is its angle of
incidence. In this two-dimensional phase space, ray optics locates
the incident and reflected rays in Fig.\ \ref{fig1} both at the
same point $\phi=\phi_0$, $\sin\chi=\sin\chi_0$, where furthermore
$\phi_0=\chi_0$ for the present case that the incident ray is
oriented into negative $x$ direction.

In wave optics, the corresponding two-dimensional Gaussian beam is
described by the wavefunction
\begin{eqnarray} &&\Psi_{\rm in}(r, \phi) = \sum_m
 c_m^- H_m^-(k r) e^{i m \phi},\\
&& c_m^- = e^{i \left(\chi_0-\phi_0-\frac\pi 2\right)
m-\frac{w^2}{2} (m - kR \sin \chi_0)^2} , \label{wavefunction}
  \end{eqnarray}
where $H_m^{\pm}$ are Hankel functions and $w$ is the width of the
beam in the polar angle $\phi$. Since we are interested in the
corrections in leading order of $kR$, we assume that the curvature
is locally constant. Then the reflected beam has the wavefunction
 \begin{eqnarray} &&\Psi_{\rm refl}(r, \phi) = \sum_m
 c_m^+ H_m^+(k r) e^{i m \phi},\\
&& c_m^+ = c_m^- \frac{ n J_m'(nkR)H_m^-(kR)-H_m^{-\prime}
(kR)J_m(nkR) }
              {H_m^{+\prime} (kR)J_m(nkR) - n J_m'(nkR) H_m^+(kR)}
,
              \quad
\label{reflected}
  \end{eqnarray}
where $J_m$ denotes the Bessel function and $n$ is the refractive
index on the other side of the interface.

In order to analyze the exact wave pattern in the phase space of
Birkhoff coordinates we utilize Husimi representations in the near
field of the dielectric interface. These Husimi functions are
obtained by overlapping the incoming and reflected beam at the
interface with a minimum uncertainty wave packet centered around
$(\phi,\sin\chi)$ \cite{epl_husimi},
\begin{eqnarray}
 &&{\cal H}^\pm
\left(\phi,\sin\chi \right) =\cos\chi\times
\nonumber\\
&&{}\times \left| \sum_m c_m^\pm H_m^\pm (kR) e^{i (m-kR\sin\chi)
\phi  - \frac{w^2}{2} (m-kR\sin\chi)^2 } \right|^2. \qquad
\label{husimis}
 \end{eqnarray}

\begin{figure}
(a) incident, near field\hspace{1cm} (b) reflected, near field\\
\includegraphics[width=0.475\columnwidth]{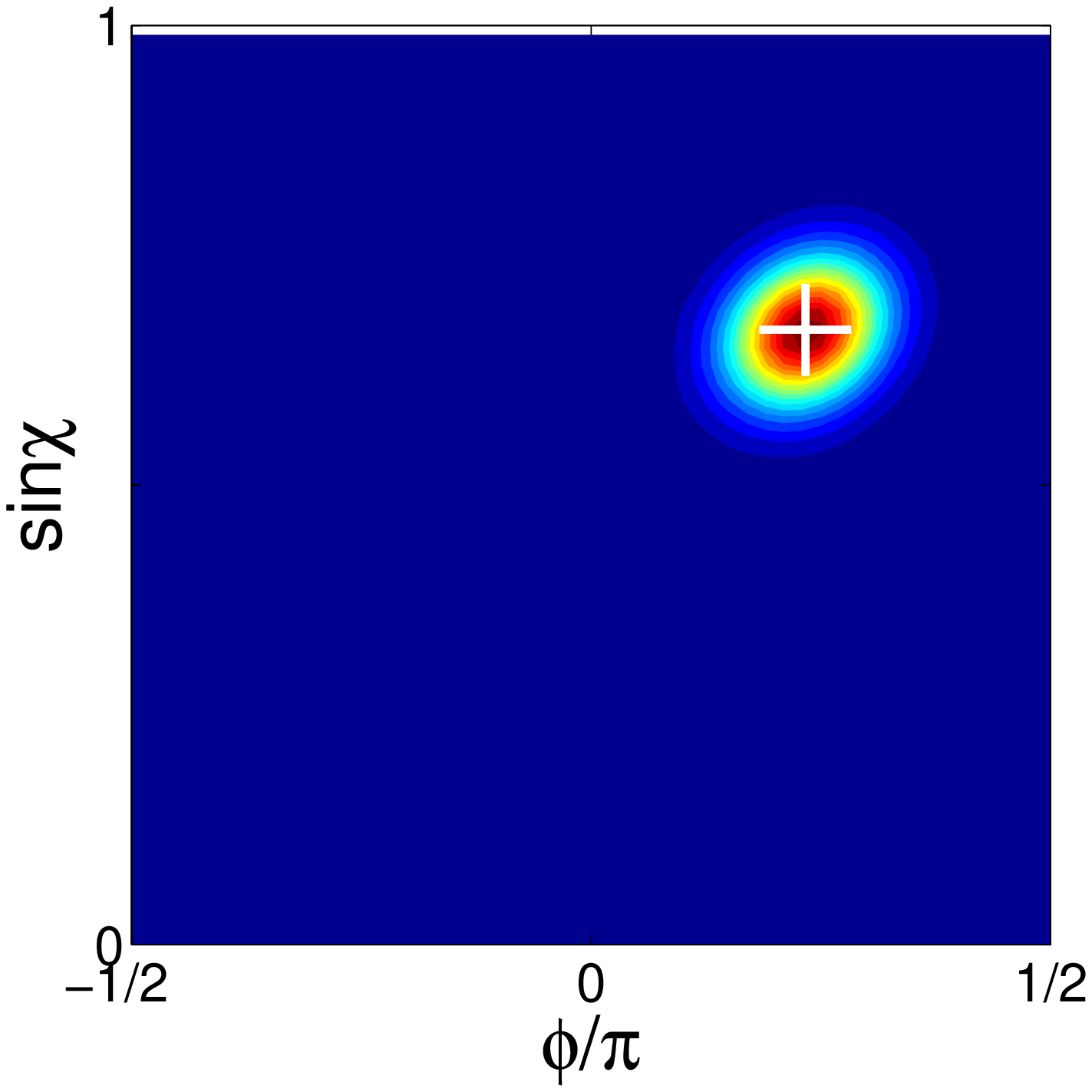}
\includegraphics[width=0.475\columnwidth]{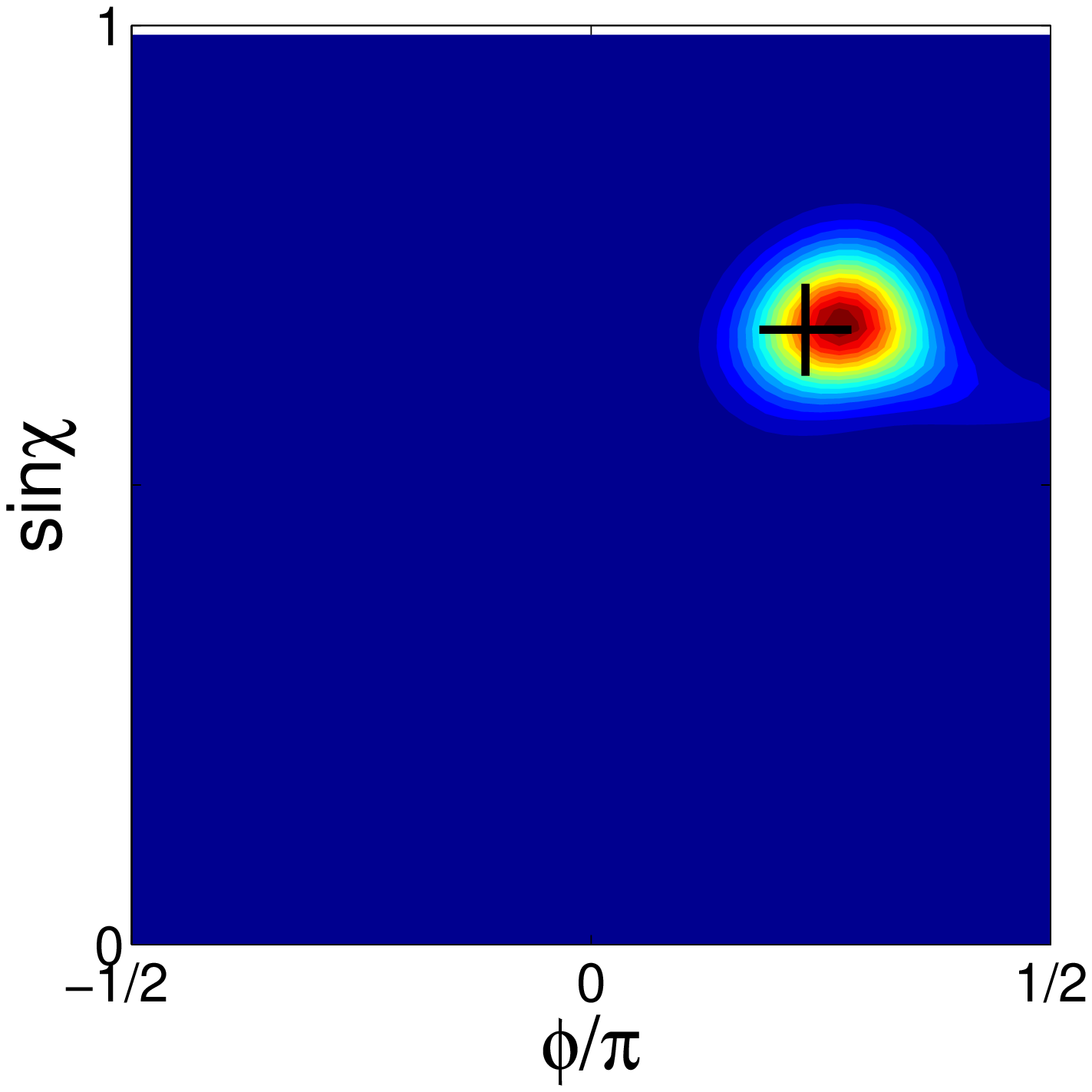}
\caption{ (Color online) Near-field Husimi phase space
representations (\ref{husimis}) of the wave pattern of Fig.\
\ref{fig1}. The left panel shows the Husimi function ${\cal H}^-$
of the incident beam, while the right panel shows the Husimi
function  ${\cal H}^+$ of the reflected beam. The crosses $+$
indicate the ray-optics prediction for the point of highest
phase-space density, which is accurate for the incident beam, but
not for the reflected beam. The displacement into $\phi$ direction
can be related the Goos-H{\"a}nchen shift, while the displacement
into $\sin\chi$ direction is the consequence of Fresnel filtering.
\label{fig2}
  }
\end{figure}

The Husimi phase-space representations of the wave pattern of
Fig.\ \ref{fig1} is shown in Fig.\ \ref{fig2}. The width $w =
\sqrt{\pi/kR}$ of the incident Gaussian beam is chosen such that
it yields an optimal approximation of a classical ray with
comparable uncertainties in the propagation direction and the
point of impact. This results in the almost-circular phase-space
density in the left panel. The location of the maximal phase-space
density corresponds well with the ray-optics prediction
$(\chi_0,\sin\chi_0)$, indicated by the cross $+$. The phase-space
representation of the reflected beam is shown in the right panel.
Clearly the position  $(\phi_{\rm max},\sin\chi_{\rm max})$ of the
maximal phase-space weight of the reflected beam is displaced from
the ray-optics prediction, as had to be expected from the
inspection of the wave pattern in Fig.\ \ref{fig1}.

The displacement in $\sin\chi$ direction can be explained by {\em
Fresnel filtering}, which was introduced by Tureci and Stone
\cite{ff,starofdavid}: A realistic beam has an uncertainty
$\gtrsim 1/(kRw)$ of its propagation direction which results in a
spreading of the angle of incidence. The angle of incidence is
further spread because of the curvature of the interface over the
focal width $Rw$. The Fresnel reflection coefficient displays an
angular dependence which favors the reflection of wave components
with a larger angle of incidence. This increases the beam's angle
of reflection, by an amount which we identify with the
displacement
\begin{equation}
\Delta\chi=\chi_{\rm max}-\chi_0 . \label{deltachi}
\end{equation}

The displacement into $\phi$ direction can be interpreted as a
Goos-H\"anchen shift (GHS), first discovered for planar interfaces
in 1947  \cite{ghs_entdeck} (for recent works see Refs.\
\cite{koreaner,ghs_phaseconj,ghs_tran,ghs_chauvat,ghs_photcryst}).
This shift arises from the penetration of the evanescent wave into
the optically thinner medium \cite{ghs_lotsch,ghs_lai}. We
identify the resulting lateral displacement  of the reflection
point along the physical interface with
\begin{equation}
\Delta\phi=\phi_{\rm max}-\phi_0 \label{deltaphi} .
\end{equation}

\begin{figure}
\includegraphics[width=\columnwidth]{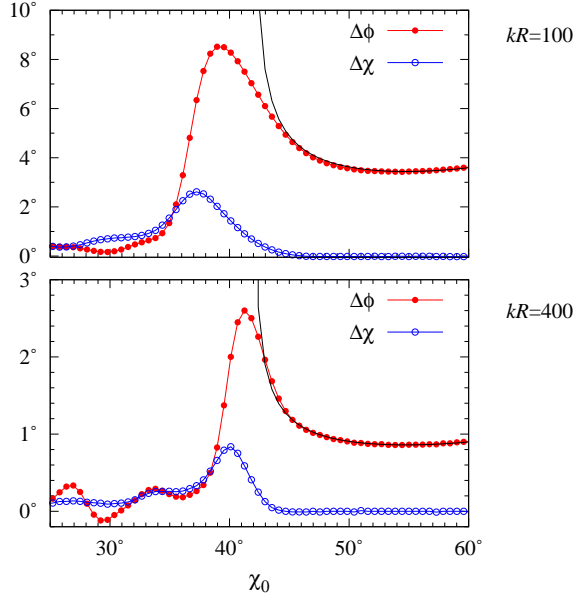}
 \caption{
(Color online) Angle-of-incidence dependence of the Goos-H\"anchen
shift $\Delta\phi$ and the Fresnel filtering correction $\Delta
\chi$ in the near field of a curved interface with $kR=100$ (top
panel) and  $kR=400$ (bottom panel). The remaining parameters are
as in Fig.\ \ref{fig1}. The black line shows the classical result
for the Goos-H\"anchen shift by Artmann \cite{ghs_artmann}.
\label{fig3}
  }
\end{figure}

The angle-of-incidence dependence of $\Delta \phi$ and $\Delta
\chi$ is shown in Fig.\ \ref{fig3}. Both corrections are most
pronounced around the critical angle of incidence $\chi'\approx
41.8^\circ$, and are sizeable effects even for rather large values
of $kR$. Beyond the critical angle,  $\Delta \phi$ approaches the
classical result for the GHS by Artmann
\cite{ghs_artmann,ghs_lotsch}, which is derived in the regime of
total reflection $\chi>\chi'$ at a planar interface.

ARO consists in propagation of the reflected beam with point of
reflection given by $\phi^{\rm ARO}=\phi_0+\Delta\phi$ and angle
of reflection given by $\chi^{\rm ARO}=\chi_0+\Delta\chi,$
resulting in a propagation direction
\begin{equation}
\theta^{\rm ARO}=\phi_0+\chi_0+\Delta\phi+\Delta\chi
\label{thetaARO}
\end{equation}
(see the right panel of Fig.\ \ref{fig1}). Note that the
corrections $\Delta\phi$ and $\Delta\chi$ both  have been
determined in the near field of the interface [see Eqs.\
(\ref{deltachi},\ref{deltaphi})]. Hence, within the idealization
of beams by rays,  ARO agrees exactly with wave optics in the near
field of the interface. The question is then whether the ARO ray
parameters deliver accurate predictions also in the far field
(where the beam may encounter another optical component or a
detector).
 Hence, we now test the accuracy of ARO by examination of its
predictions for the far-field radiation characteristics.

\begin{figure}
(a) incident, far field\hspace{1cm} (b) reflected, far
field\\
\includegraphics[width=0.475\columnwidth]{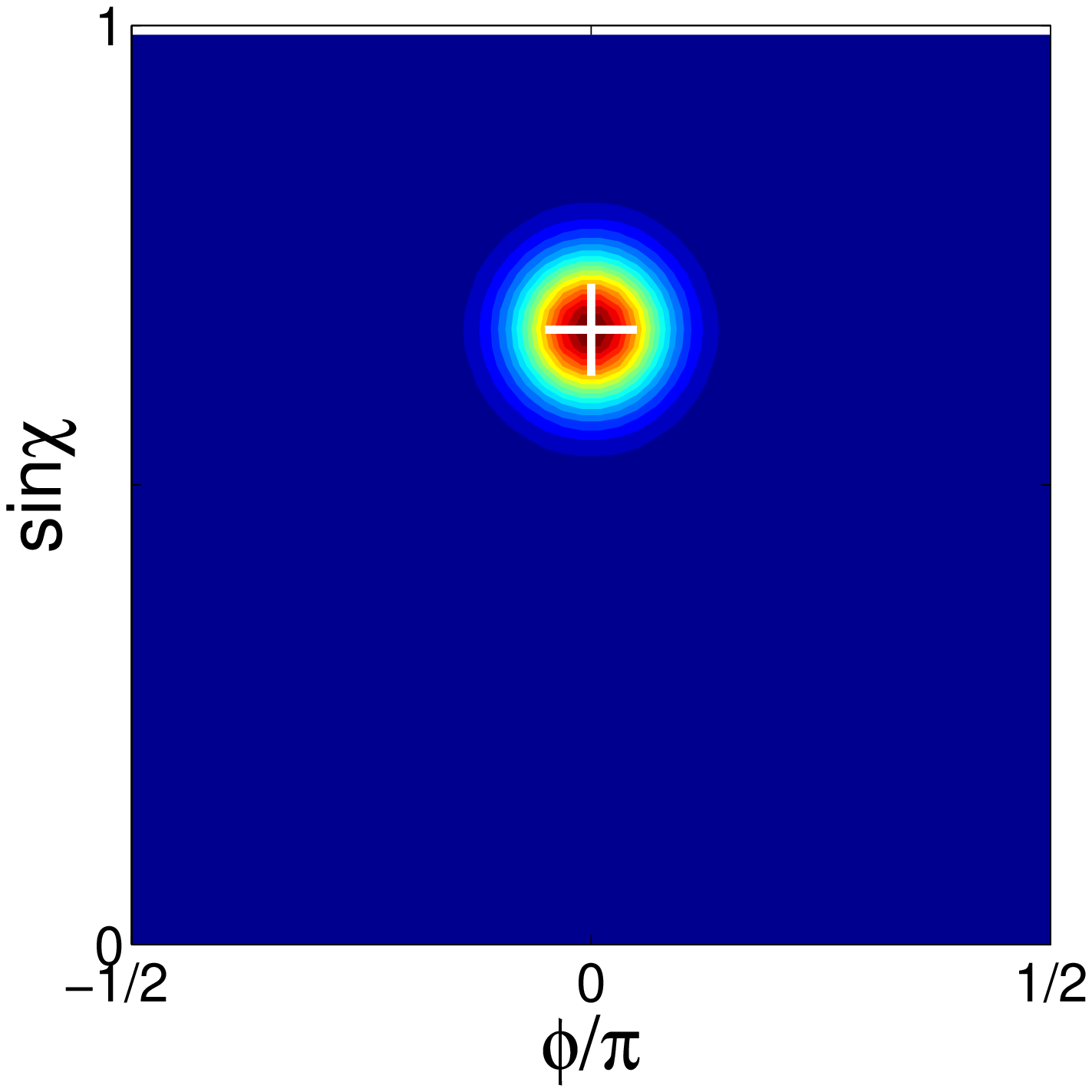}
\includegraphics[width=0.475\columnwidth]{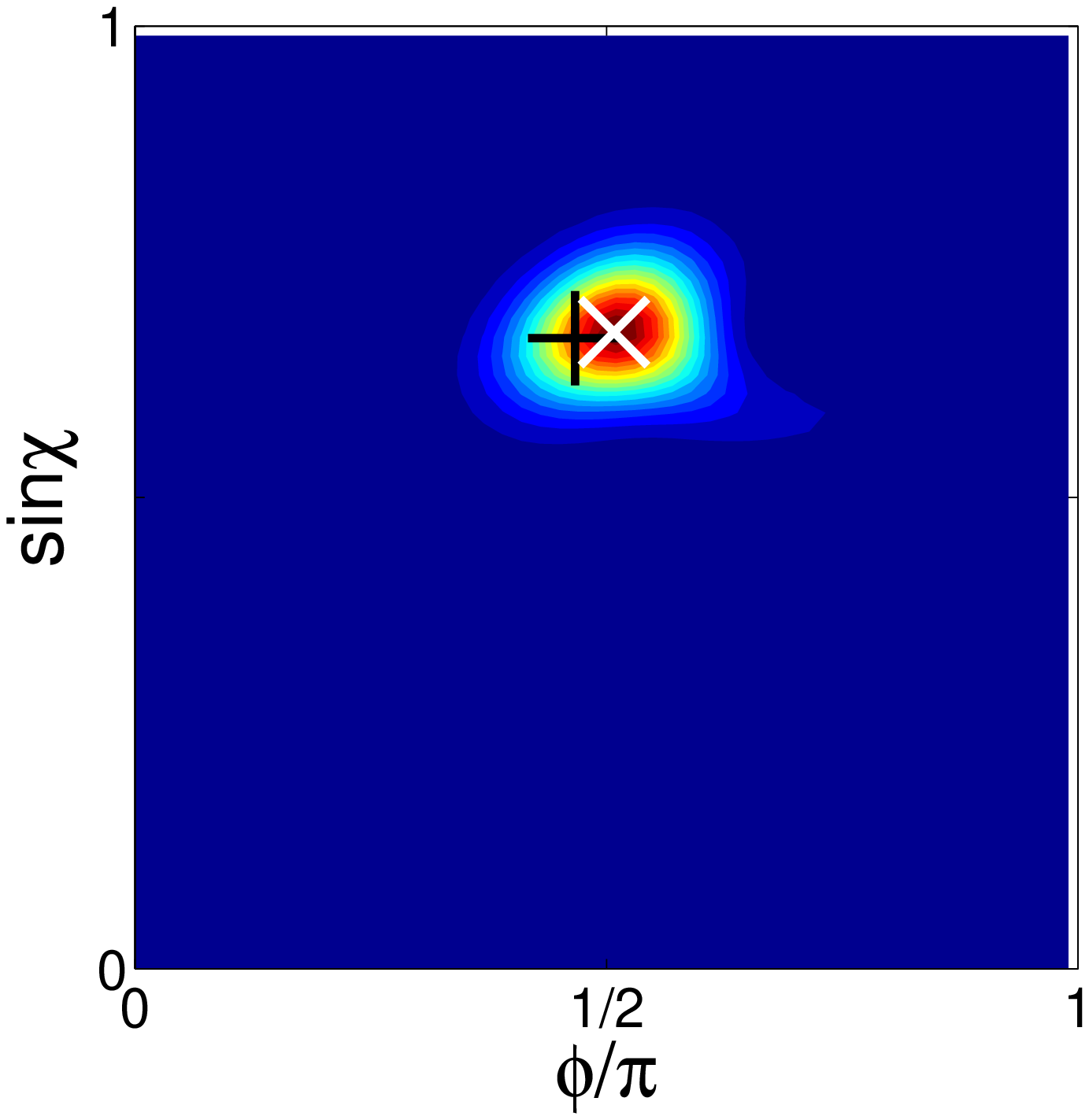}
\caption{ (Color online) Same as Fig.\ \ref{fig2}, but for the far
field, where the Husimi representations $\tilde{\cal H}^\pm$ are
given by Eq.\ (\ref{husimifar}). The diagonal cross $\times$
indicates the ARO prediction for the reflected beam, which
incorporates the Goos-H{\"a}nchen shift and Fresnel filtering.
\label{fig4}
  }
\end{figure}

Figure \ref{fig4} assesses these predictions for the wave pattern
of Fig.\ \ref{fig1} by means of  Husimi phase-space representations
of the incident and reflected beam in the far field of the
interface,
\begin{eqnarray}
 &&\tilde{\cal H}^\pm \left(\phi,\sin\chi \right)=
\nonumber
\\
&&
\left| \sum_m c_m^\pm e^{i (m-kR\sin\chi) \phi \mp i\pi m/2 -
\frac{w^2}{2} (m-kR\sin\chi)^2 } \right|^2 .\qquad \label{husimifar}
 \end{eqnarray}
In the far field,  the phase-space coordinate $\phi$ coincides
with the propagation direction $\theta$, while $\sin\chi$ is still
related to the impact parameter  $R\sin\chi$ (this coordinate is
preserved because of angular-momentum conservation with respect to
the center of the circle of curvature). The incident beam
propagating into negative-$x$ direction is thus represented by phase
space coordinates $(\phi,\sin\chi)= (0, \sin\chi_0)$. Ray-optics
predicts that the reflected beam has phase-space coordinates
 $(\phi,\sin\chi)= (\theta^{\rm RO},\sin\chi_0)$, where
$\theta^{\rm RO}=\phi_0+\chi_0,$ while ARO predicts that the
reflected beam is located at $(\phi,\sin\chi)=(\theta^{\rm
ARO},\sin\chi^{\rm ARO}$). The position  $(\tilde\phi_{\rm
max},\sin\tilde\chi_{\rm max})$ of the maximal phase-space density
of the reflected beam in the far field (right panel of Fig.\
\ref{fig4}) indeed corresponds well to the ARO prediction
(indicated by $\times$), but deviates distinctively from the
ray-optics prediction (indicated by $+$).

\begin{figure}
\includegraphics[width=\columnwidth]{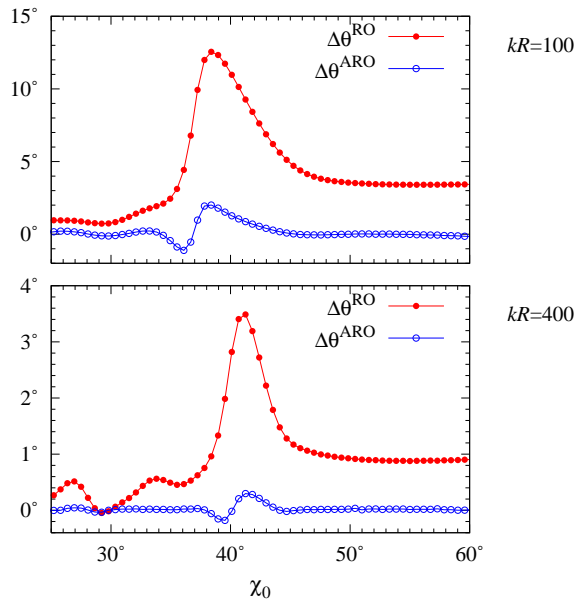}
 \caption{(Color online)
Angle-of-incidence dependence of the deviation of the far-field
radiation direction $\theta$ from the predictions of ray optics
(RO) and amended ray optics (ARO). Top panel: $kR=100$. Bottom
panel: $kR=400$. The remaining parameters are as specified in
Fig.\ \ref{fig1}.
 \label{fig5}
  }
\end{figure}

In Fig.~\ref{fig5}  the far-field radiation direction $\theta$ is
analyzed as a function of the angle of incidence. One of the
curves is the deviation $\Delta\theta^{\rm RO}=\tilde\phi_{\rm
max}-\theta^{\rm RO}$ of the observed radiation direction from the
prediction of conventional ray optics. For $kR=100$, the maximal
deviation is $\approx  12.5^\circ$ and occurs about $4^\circ$
below the critical angle of incidence. The plot also shows the
deviation  of ARO, $\Delta\theta^{\rm ARO}=\Delta\theta^{\rm
RO}-\Delta\phi-\Delta\chi$ . It is seen that ARO improves the
agreement to $2^{\circ}$ close to the critical angle and agrees
even better away from it. For larger size parameters $kR=400$, the
maximal disagreement between ray optics and wave optics drops to
$\approx 3.5^\circ$ and occurs at about $1^\circ$ below the
critical angle of incidence. The ARO prediction agrees within
$0.3^\circ$ around the critical angle, and the agreement is almost
perfect away from it.

To summarize, we developed a systematically amended ray-optics
description of the reflection of Gaussian beams from the curved
dielectric interfaces of optical microresonators. This description
incorporates the Goos-H\"anchen shift of the reflection point
along the interface and the Fresnel-filtering enhancement of the
angle of reflection. The corrections were determined by analysis
of exact wave-optical beams in a phase space where one coordinate
is associated with the point of incidence or reflection along the
interface, while the other one is related to the angle of
incidence or reflection, respectively. Fresnel filtering and the
Goos-H\"anchen effect displace the reflected beam along
independent phase-space directions. Hence, these displacements in
principle exhaust all possibilities of amending ray optics while
still keeping the basic assumption of propagation along straight
lines in optically homogeneous media.

Amended ray optics is applicable to microresonators with smooth
boundaries where the dimensionless radius of curvature $kR$ is
large, which is realized in most experiments. This includes the
popular examples of multipole deformations
\cite{bowtie,starofdavid,sangwookkim,nature,optlett_interf,pre_annbill,schwefel}
or stadium geometries \cite{japaner,lebental,morelasers}.
Complementary techniques exist to describe the diffraction of
beams at sharp corners where formally $kR=0$ \cite{diffraction};
see Ref.\ \cite{jan} for an application to hexagonally shaped
resonators \cite{VKLISLA98}. It remains to be seen whether both
techniques can be interlaced to describe geometries which combine
both curved interfaces and sharp corners
\cite{morelasers2,koreaner}; moreover, whether both techniques can
be unified in the challenging regime of a local curvature with
$kR=O(1)$.

We thank S.-Y.~Lee, O.~Legrand, H.~Schwefel, R.~Weaver, and
J.~Wiersig for discussions. This work was supported by the
Alexander von Humboldt Foundation and the European Commission,
Marie Curie Excellence Grant MEXT-CT-2005-023778
(Nanoelectrophotonics).

\end{document}